%% file: level-spectroscopy.tex
\documentclass[12pt]{article}
 \usepackage{fr-jp}
 \usepackage[dvips]{graphicx}

 \title{\bf BKT transition and level spectroscopy}
 \author{Kiyohide Nomura$^{1}$\thanks{knomura@stat.phys.kyushu-u.ac.jp}
 and Atsuhiro Kitazawa$^2$\thanks{kitazawa@riam.kyushu-u.ac.jp}\\
 \em $^{1}$Department of Physics, Kyushu University, Fukuoka 812-8581, JAPAN \\
 \em $^{2}$Research Institute for Applied Mechanics, Kyushu University \\
 \em Kasuga, Fukuoka 816-8580, JAPAN}
 \date{}

\begin{document}
 \maketitle
 \begin{abstract}
 Berezinskii-Kosterlitz-Thouless (BKT) transition is 
 one of the instability mechanisms for the Tomonaga-Luttinger liquid. 
 But in the BKT transition, there are logarithmic-correction problems, 
 which make it very difficult to treat BKT transitions numerically. 
 We propose a method, ``level spectroscopy'',
 to overcome such difficulties, based on the renormalization group
 analysis and the symmetry consideration. 
 \end{abstract}

 \section{Introduction}

 Tomonaga-Luttinger (TL) liquid \cite{Tomonaga,Haldane} 
 is an important concept for
 one dimensional (1D) quantum systems (spin, electron systems, nanotube, 
 etc. at $T=0$) with the continuous symmetry (U(1),SU(2), etc.),
 and TL liquid is closely related to several 2D classical models 
 (6-vertex model, classical XY spin,
 superconducting film, roughening transition, etc. at $T \neq 0$).  
 Therefore, it becomes crucial to understand instabilities of TL 
 liquid.

 One of the instabilities of TL liquid is
the Berezinskii-Kosterlitz-Thouless (BKT) transition 
 \cite{Berezinskii, Kosterlitz-Thouless,Kosterlitz}.  
 Although the BKT transition has been well known, 
 it has not been recognized that there are pathological problems to analyze 
 numerically BKT transitions.  One of these problems is that the finite
 size scaling technique \cite{Barber}, 
 which is successful for second order transitions, 
 can not be applied for BKT transitions \cite{Solyom-Ziman,Edwards},
 since there are logarithmic corrections from the marginal coupling.  
 Recently, combining the renormalization group calculation and the symmetry, 
 we have developed a remedy, ``level spectroscopy'',
 to overcome these difficulties. 

 In the next section, we compare the BKT transition with the second order
 transition.  
 In \S 3, we introduce the concept of the level spectroscopy and how to
 use it. 
 In \S 4, we deduce the level spectroscopy from the sine-Gordon model,
 which is an effective model to describe the BKT transition. 
 For readers who are not familiar with the sine-Gordon model,
 we recommend to read \S 5 (physical examples) before \S 4.

 \section{BKT transition versus second order transition}

First, we briefly review the renormalization concept.  
Let us consider a $d$-dimensional classical Hamiltonian 
 (or an action for a $(d-1)+1$ dimensional quantum system) \cite{Cardy}
 \begin{equation}
  H = H_0 + \sum_{j} y_j \int \psi_j (x) \frac{d^d x}{\alpha^d},
   \label{eq:scaling-Hamiltonian}
 \end{equation}
 where $H_0$ is scale invariant, $\psi_j (x)$ is a local order parameter, 
 $y_j$ is an effective coupling constant (or some external field),
 $\alpha$ is a short-range cutoff. 

 When changing a scale as $ \alpha' = b\alpha$, 
 effective local order parameters  change as 
 $\psi'_j =b^{x_j}\psi_j$ ($x_j$: scaling dimension)
 and according to this, effective couplings change as $y'_j=
 b^{d-x_j} y_j$. If $y_j=0$, then $y'_j$ remain 0 (fixed point).  
 We call the case $y'_j$ diverging for $b \rightarrow \infty$ 
(i.e. $x_{j} < d$) as relevant,
 whereas  the case $y'_j$ converging (i.e. $x_{j} > d$) as irrelevant.

 At the fixed point, the system is scale invariant, therefore the correlation
 length is infinite $\xi=\infty$. 
 Although regions which flow into the fixed point are not strictly scale
 invariant, $\xi=\infty$ in these regions (critical regions).

 For the infinitesimal scale transformation 
 $\alpha' = \alpha e^{d l} \approx \alpha(1+ d l) $, 
 one can treat eq. (\ref{eq:scaling-Hamiltonian}) perturbatively. 

 \subsection{Second order transition}

\begin{figure}[here]
  \begin{center}
  \includegraphics[width=5cm]{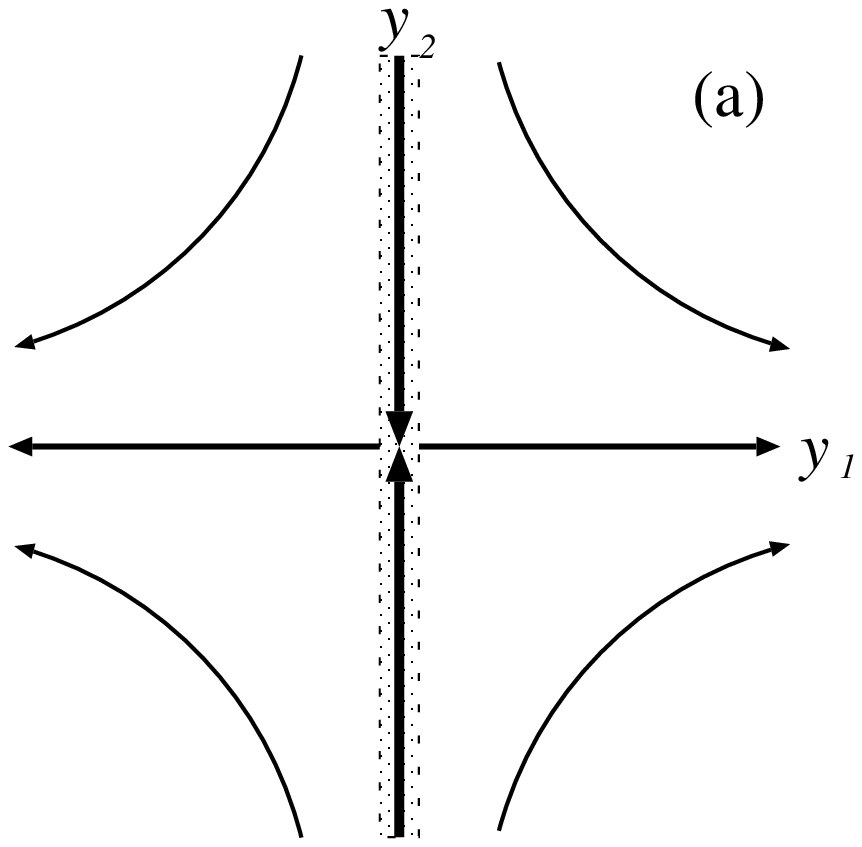}
  \hskip 1cm
  \includegraphics[width=5cm]{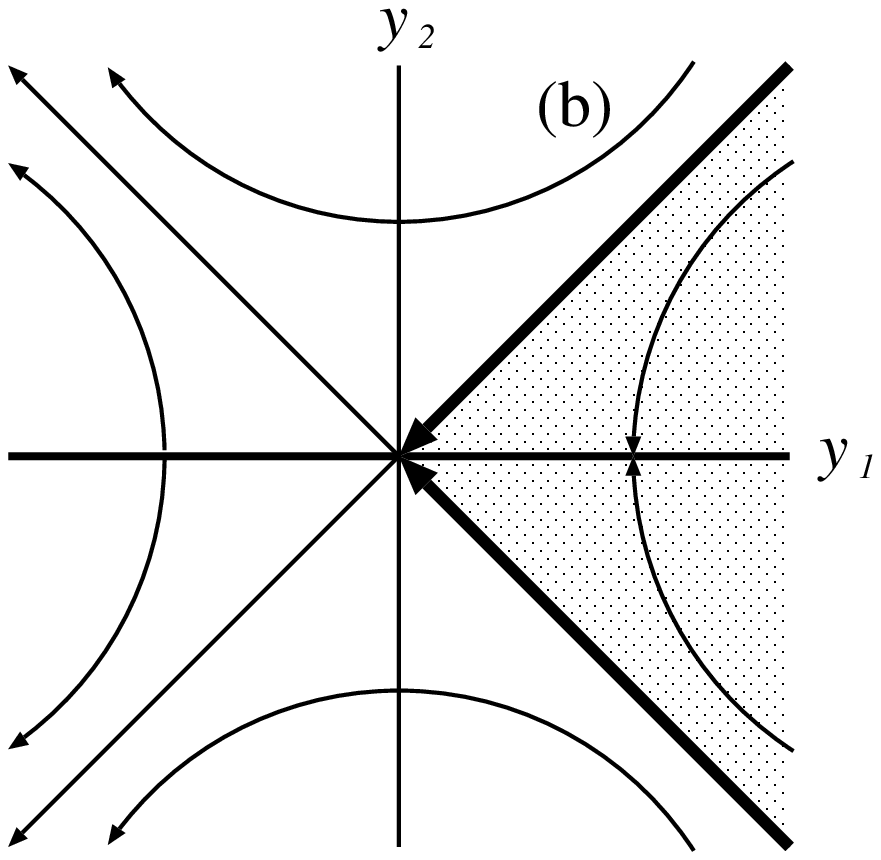}
  \caption{Renormalization flows for (a) second order transition, (b) BKT
  transition.
  Shaded parts are critical region.}
  \label{fig:renormalization-flow}
   \end{center}
 \end{figure}

 The scaling equations for second order transitions are
 \begin{equation}
  \frac{d y_{j}(l)}{dl} = (d - x_{j}) y_{j}(l) \equiv \beta_j (\{
  y_{j}(l)\}) \qquad  (l \equiv \ln(L), L:{ \rm system \; size}).
 \end{equation}
 As an example, let us consider the two scaling field case. 
 When $\psi_1$ is a relevant operator and  $\psi_2$ is an
 irrelevant one,
 the renormalization group flow is given in
 Fig. \ref{fig:renormalization-flow} a.

 Since the second order transition occurs where the sign of the $\beta$ function
 (for a relevant coupling $y_1$) changes, one can use this fact to determine the
 critical point \cite{Barber}.

 \subsection{BKT transition}

The BKT transition ($d=2$) is described by the following RG equation \cite{Kosterlitz}
 \begin{equation}
 \fbox{$\displaystyle
  \frac{d y_{1}(l)}{dl} = - y_{2}(l)^2, \;
  \frac{d y_{2}(l)}{dl} = - y_{1}(l) y_{2}(l)
  $},
  \label{eq:BKT-renormalization}
 \end{equation}
 where fixed points are $y_2=0$.  
 All of the points in $|y_2| < y_1$ are renormalized to $y_2=0$,
 thus in this region correlation lengths are infinite ($\xi=\infty$) or
 massless. Other regions are massive ($\xi$ finite), except $y_2=0$
 (see Fig \ref{fig:renormalization-flow} b). 
 We call the region within $|y_2| < y_1$ as the massless region,  
 the lines $y_2 = \pm y_1$ as the 
 BKT critical lines, $y_1=0$ as the Gaussian fixed line.
 The point $y_{1}=y_{2}=0$ has a special meaning (two BKT lines and one
 Gaussian line intersect), and we call it as the BKT multicritical point. 

 Note that on the BKT line, the coupling $y_{1}$ is marginal
 (i.e. $x_{1}=2$) and it behaves
 as $y_{1}(l)= 1/l= 1/ \ln (L)$.  Another important fact is that 
 on the Gaussian fixed line ($y_{2}=0$), the scaling dimension of $\psi_{2}$
 is varying with $y_{1}$.  

 \subsection{Comparison between  second order and BKT transition}
 \begin{enumerate}
  \item Critical region ($\xi=\infty$) is isolated in the second order
	transition, whereas it is extended in the BKT transition.

  \item Zero point of the $\beta$ function corresponds to the critical point in
	the second order transition,
	whereas for the BKT, it has no special meaning.
  \item Since the renormalization group behavior on BKT critical lines ($y_2=\pm
	y_1$) is marginal, there appear logarithmic corrections
	$1 /\ln (L)$, {\it e.g.}, in the correlation function as 
	\cite{Kosterlitz}
 \begin{equation}
	r^{1/4}(\ln r)^{1/8},
 \end{equation}
	or in the energy gap for the finite size system as \cite{Cardy86,Affleck-Gepner-Schulz-Ziman}
 \[
	\Delta E (L) = \frac{2 \pi v}{L} \frac{1}{8}
	\left( 1 -  \frac{1}{2}\frac{1}{\ln L} \right),
 \]
	therefore finite size effects are very
	large. 
 \end{enumerate}

 All of them make it very difficult to treat the BKT transition numerically  
 (N.B. universal jump is also affected by logarithmic corrections, but
 O($1/(\ln L)^2$)).

 \section{Level spectroscopy}

 The BKT transition occurs where some physical quantities change from
 irrelevant to relevant. Thus it is useful to investigate scaling
 dimensions near marginal. In fact, on the Gaussian line (i.e.,
 without interaction), several 
 scaling dimensions cross at the BKT multicritical point (see
 Fig. \ref{fig:scaling} a, or
 eqs. (\ref{eq:scaling-dim}), (\ref{eq:scaling-dim2})).

 Based on the conformal field theory (CFT) in 2D \cite{CFT}, one can obtain the scaling
 dimensions using the energy gap for the finite system.
 The relation between the energy gap $\Delta E_j$ of the 
 system size $L$ and the scaling dimension $x_j$ is \cite{Cardy,Cardy84}
 \begin{equation}
  \fbox{$\displaystyle
  \Delta E_j = \frac{2 \pi v x_j}{L}
  $},
   \label{eq:scaling}
 \end{equation}
where $v$ corresponds with a renormalized Fermi velocity, or a spin wave velocity.

 Each excitation can be classified by quantum numbers
 ($m$ (magnetization or electron density, related with U(1) symmetry), $P$
 (parity), $q$ (wave number)).  

 In the normal BKT transition, there is no symmetry breaking in
the massive phase.  But in general, the BKT transition may be combined with
a discrete symmetry, and there occurs a $Z_{p}$ discrete symmetry breaking in
the massive phase.

 \bigskip

 Procedure to use level spectroscopy
 \begin{enumerate}
  \item Classification of BKT transitions with the discrete symmetry
	\begin{enumerate}
	 \item BKT transition without symmetry breaking (\S 3.1, \S 3.3)

	       ({\it e.g.} 2D classical XY, integer $S$ XXZ quantum spin chain )

	       From  table \ref{table:table1},
	       choose the excitation with some quantum numbers,
	       then from the level crossing, determine 
	       BKT transition line.

	 \item BKT transition with the $Z_2$ symmetry breaking (\S 3.2)

	       ({\it e.g.} 6 vertex model, half-integer $S$ XXZ quantum
	       spin chain )

	       From  table \ref{table:table2},
	       choose the excitation with some quantum numbers,
	       then from the level crossing, determine 
	       BKT transition line.

	 \item BKT transition with the $Z_p$ ($p>2$ integer) symmetry
	       breaking 

	       ({\it e.g.} 2D $p$-state clock model)

	       Although we do not explain the $Z_p$ case here, 
	       this case has been discussed in \cite{Nomura}. Note that there
	       is a difference for $p$ even or odd case.

	\end{enumerate}
  \item Checking of the universality class (elimination of logarithmic corrections)
	\begin{enumerate}
	 \item Scaling dimension (\S 3.4)

	       From tables \ref{table:table1}, \ref{table:table2},
	       choose the excitations eliminating logarithmic corrections
	       each other,
	       and check the universality class.

	 \item Central charge $c=1$ (\S 3.5)

	\end{enumerate}
 \end{enumerate}

 \subsection{Normal BKT transition}

\begin{figure}[here]
  \begin{center}
  \includegraphics[width=5cm]{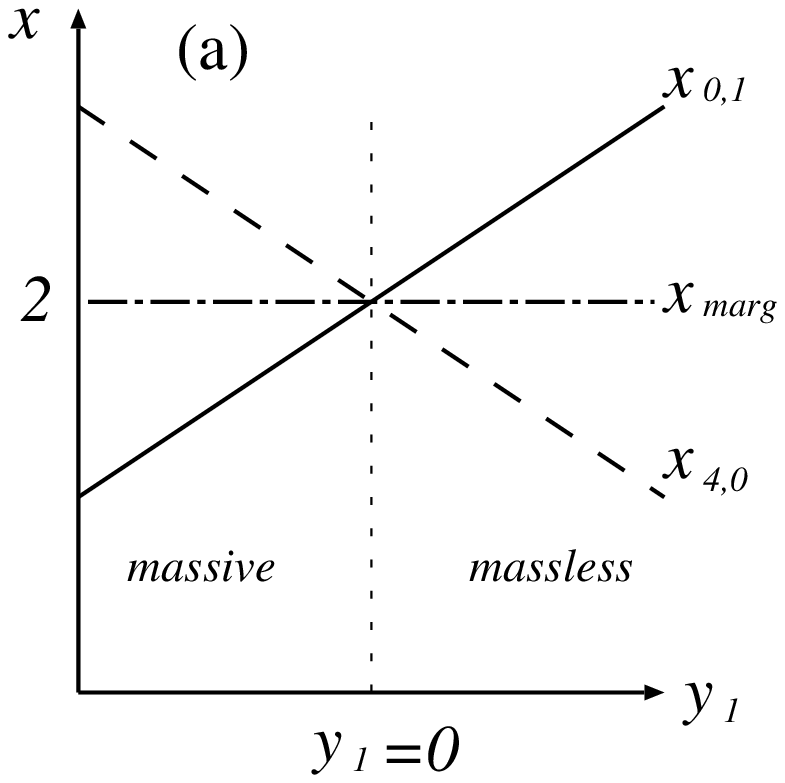}
  \hskip 1cm
  \includegraphics[width=5cm]{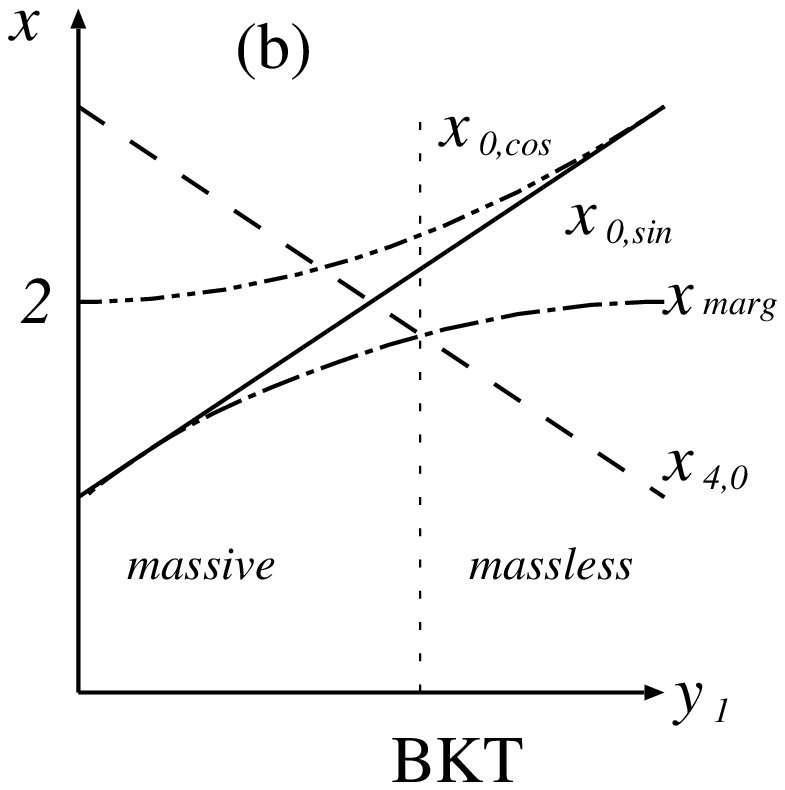}
  \caption{
  Scaling dimension  $x$ in the neighborhood of BKT transition
  (a) $y_2=0$ (on the Gaussian fixed line) 
  (b) $y_2 \neq 0 $.
  }
  \label{fig:scaling}
\end{center}
 \end{figure}

 In the normal BKT transition, there is no symmetry breaking.
 In table \ref{table:table1}, we show the relation between the scaling
 dimension (or excitation (\ref{eq:scaling})) and quantum numbers
 \cite{Nomura}.
 In the neighborhood of the scaling dimension $x=2$,
 on the BKT line $y_2 (l) = \pm y_1 (l)$,
 there is a level crossing of excitations with quantum numbers ($m=\pm
 4,P=1,q=0$) and ($m=0,P=1,q=0$) 
 (see table \ref{table:table1}, Fig. \ref{fig:scaling}),
 thus we can determine the BKT transition line.

 Next, since the ratios of logarithmic corrections ($1/\ln L$) on the BKT
 transition line are $2:1:-1$ (corresponding to
 $x_{0,cos},x_{0,sin},x_{\pm 4,0}$
 in table \ref{table:table1}),
 we can eliminate logarithmic corrections and check the universality class.

 This level crossing on the BKT transition reflects the (hidden) SU(2)
 symmetry \cite{Halpern}, thus it is correct up to higher order loops. 
 We can see this SU(2) symmetry explicitly, including the twisted boundary
 condition (see \S 3.3).
 \begin{table}[here]
\begin{center}
  \input{table1}
  \caption{
  Renormalized scaling dimensions $x$ and quantum numbers ($m,P,q$)
  for the normal BKT transition.
  PBC denotes periodic boundary condition, TBC twisted boundary condition.
  $t$ is a distance from the BKT critical line,
  defined as $y_2(l)= \pm y_1(l)(1+t)$.  
  Couplings $y_1,y_2$ follow renormalization group equations
  (\ref{eq:BKT-renormalization}), and on the BKT transition line
  $y_1(l) = \pm y_2(l) = 1/ \ln (L/L_0)$.
  }
  \label{table:table1}
\end{center}
 \end{table}

 \subsection{BKT transition with $Z_2$ symmetry breaking}

 Next we consider the BKT transition coupled with a discrete symmetry.
 Especially in the BKT transition with the $Z_2$ symmetry breaking 
 (in the massive region), the level crossing of the lowest excitations in
 each region corresponds to the phase boundary. 

\begin{figure}[here]
\begin{center}
  \includegraphics[width=5cm]{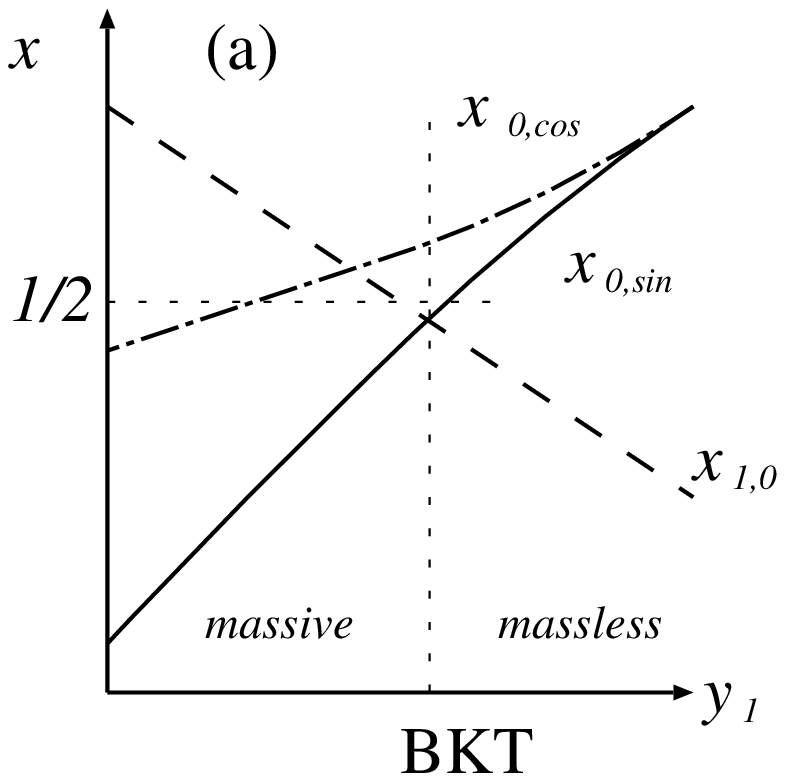}
  \hskip 1cm
  \includegraphics[width=5cm]{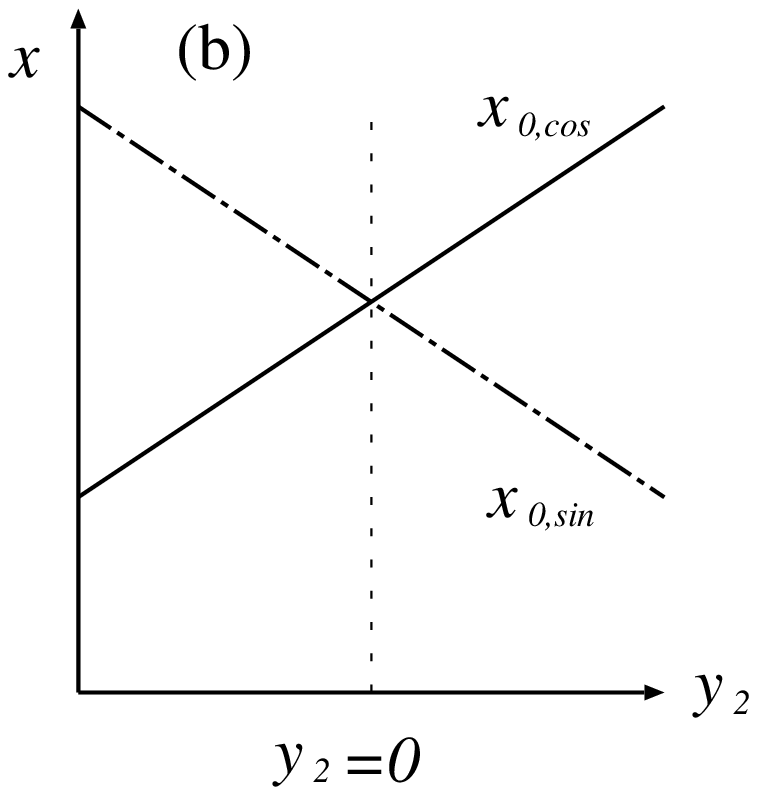}
  \caption{
  Scaling dimensions $x$ for BKT transition with $Z_2$ symmetry.
  (a) Near the BKT transition line ($y_2>0$ case) (b) Near the Gaussian
  fixed line.
  }
  \label{fig:scaling4}
\end{center}
 \end{figure}

 We introduce a quantum number corresponding to the $Z_2$ symmetry
 (wave number  $q=0,\pi$ etc.).  
 In table \ref{table:table2} we summarize quantum numbers and excitations
 \cite{Giamarchi-Schulz,Nomura-Okamoto}.
 Level crossings on BKT transition lines can be observed not only 
 $x=2$ (N. B. quantum numbers differ from table \ref{table:table1}),
 but also $x=1/2$, 
 where the excitations with quantum numbers ($m=0,q=\pi,P= \pm 1$) and
 ($m= \pm 1$) show a level crossing
 (Fig. \ref{fig:scaling4} a, table \ref{table:table2}).

 About the Gaussian fixed line $y_2=0$,
 each of the $Z_2$ symmetry broken massive phases has different parity, 
 thus it occurs a level crossing of excitations $m=0,q=\pi,P= \pm 1$
 (Fig. \ref{fig:scaling4} b).  

 \begin{table}[here]
\begin{center}
  \input{table2}
  \caption{
  Renormalized scaling dimensions $x$ and quantum numbers ($m,P,q$)
  for BKT transition with $Z_2$ symmetry.
  }
  \label{table:table2}
\end{center}
 \end{table}

 \subsection{Twisted boundary condition (TBC)}

 In the previous section case, one can distinguish two massive phases
 on the both side of the Gaussian fixed line, by using $Z_{2}$ symmetry. 
 How to distinguish two massive phases 
 in the normal BKT transition?
 Using the twisted boundary condition (TBC) method \cite{Kitazawa,Kitazawa-Nomura},
 we can clarify the hidden $Z_2 \times Z_2$ symmetry. 

 TBC is expressed by the sine-Gordon language as (see \S 4)
 \begin{equation}
 \exp ( \pm i \sqrt{2} \theta(z_{0},z_{1}+L))
 =-\exp ( \pm i \sqrt{2} \theta(z_{0},z_{1})),
 \end{equation}
 or in the quantum spin language (S=1 case, see \S 5)
 \begin{equation}
 S^{x,y}_{L+j} \equiv -S^{x,y}_j, S^z_{L+j} \equiv S^z_j.
 \end{equation}

 We can also define the discrete (inversion) symmetry 
$P^{*}: S_j \leftrightarrow S_{L+1-j}$ under TBC 
(N. B. differs from the parity under PBC). Corresponding to this quantum number, 
 we can observe the level crossing of $m=0$ states under TBC at the Gaussian fixed line
 $y_{2}=0$ (see table \ref{table:table1}).

 Besides, using the TBC method, we can clarify the hidden SU(2) symmetry
 on the BKT critical line
 \cite{Nomura-Kitazawa,Halpern}. 
 That is, we can determine BKT line by the level crossing 
 between the excitations under PBC 
 $q=0,m=\pm 2$ 
 and that under TBC $m=0$.
 (see table \ref{table:table1}).

 \subsection{Universality class (scaling dimension)}

 Finally, in order to check the consistency of our method, we should
 eliminate logarithmic corrections from scaling dimensions (critical
 indexes). 
 There are several methods to eliminate logarithmic size corrections on BKT
 lines. 

 For the BKT transition with $Z_2$, relations
 \begin{eqnarray}
  (x_{0,sin}+ x_{0,cos})\times x_{\pm 1,0} =1/2, \\
  x_{\pm 2,0}/ x_{\pm 1,0} =4,
 \end{eqnarray}
 are correct up to $O(1/(\ln L )^2)$,
 and they also apply all over the critical region. 

 Similarly, for the normal BKT case, combining TBC, we can check the
 universality class as the above method \cite{Kitazawa-Nomura}.

 \subsection{Central charge}

 The BKT critical region (massless region $\xi=\infty$) also can be characterized 
 by the central charge $c=1$.
 Numerically the central charge $c$ is obtained from the ground state
 energy for the finite system as \cite{Blote-Cardy-Nightingale}
 \begin{equation}
  E_g(L) = e_g L -\frac{\pi v c}{6 L}.
   \label{eq:central-charge}
 \end{equation}
 (N. B.  the central charge is also obtained experimentally from the specific
 heat \cite{Affleck}.)

 Although there are logarithmic corrections in the effective central
 charge $c$ obtained from
 eq. (\ref{eq:central-charge}),
 they are small enough $O(1/(\ln L)^3)$ \cite{Cardy86}, thus we can
 neglect them. 

 Note that the effective central charge, obtained as
 eq. (\ref{eq:central-charge}),
 changes rapidly from $c=1$ (massless) to $c=0$ (massive)
 \cite{Inoue-Nomura,Inoue}.

 \section{Sine-Gordon model}

 There are several effective models to describe the BKT transition.
 \begin{enumerate}
  \item 
	One of the effective models for BKT transitions is 
	a sine-Gordon model, 
 \begin{equation}
   Z = \int {\cal D \phi} \exp( - \int d^2 x {\cal L}), \qquad
   {\cal L} = \frac{1}{2 \pi K} (\nabla \phi)^2
    + \frac{y_2}{2 \pi \alpha^2}\cos  p \sqrt {2}\phi,
    \label{eq:sine-Gordon}
 \end{equation}
 where $p$ is integer, $K$ is related with $y_1 = 2(K p^2/4-1)$ in
 eq. (\ref{eq:BKT-renormalization}). 
In addition to eq. (\ref{eq:sine-Gordon}), we require following features:
	\begin{enumerate}
	 \item Compactification:
	       $\phi \equiv \phi + \sqrt{2}\pi.$
	 \item 
 Canonical field $\theta$ to $\phi$
 \begin{equation}
  \partial_x \phi = - \partial_y( i K \theta), \;
   \partial_y \phi =  \partial_x( i K \theta),
 \end{equation}
	 \item  Compactification of $\theta$: 
	       $\theta \equiv \theta + \sqrt{2}\pi.$

	       This represents the continuous U(1) symmetry, 
	       because eq. (\ref{eq:sine-Gordon}) is invariant under 
		$\theta \rightarrow \theta + const$.

	 \item Discrete symmetry

	       Eq. (\ref{eq:sine-Gordon}) is invariant under the change
	       $ \phi \rightarrow \phi + \sqrt{2}\pi/p$, which corresponds to 
	       the discrete $Z_{p}$ symmetry.

	\end{enumerate}

  \item Free case ($y_2=0$) \cite{Kadanoff-Brown}

	\begin{enumerate}

	 \item 
 Correlation functions for $\phi$ and $\theta$

It is convenient to describe coordinates in 2D as complex variables: 
$z  \equiv x + i y $. Then, correlation functions for $\phi,\theta$ are 
 \begin{eqnarray}
   \langle  \phi (z) \phi(0) \rangle &=& -\frac{K}{2} \ln \left( \frac{|z|}{\alpha} \right), \;
   \langle  \theta (z) \theta(0) \rangle = -\frac{1}{2K} \ln \left(
   \frac{|z|}{\alpha}\right), \nonumber \\
   \langle  \phi (z) \theta(0) \rangle &=& -\frac{i}{2} \arg z.
 \end{eqnarray}

	 \item Vertex operators
	       \begin{equation}
		O_{m,n} \equiv \exp ( i \sqrt{2} m \theta)\exp ( i \sqrt{2} n \phi)
		 \qquad (m,n: {\rm integer}).
		 \label{eq:vertex-operator}
	       \end{equation}

	 \item Marginal operators
	       \begin{equation}
   O_{marg} \equiv (\partial_x \phi)^2 + (\partial_y \phi)^2.
	       \end{equation}

	 \item Correlations and scaling dimensions for vertex operators
	       \begin{equation}
   \langle O_{m,n}(z)O_{-m,-n}(0) \rangle 
    = \exp \left[ - 2 x_{m,n} \ln \left(\frac{|z|}{\alpha} \right) -  2
    i l_{m,n} (\arg z + \pi/2) \right].
	       \end{equation}
	       Therefore, scaling dimensions are given by
	       \begin{equation}
		\fbox{$\displaystyle x_{m,n} = \frac{1}{2} \left( \frac{m^2}{K} + n^2 K \right), \;
   l_{m,n} =mn $},
   \label{eq:scaling-dim}
	       \end{equation}
	       thus $x_{m,n}$ are varying with coupling $K$, whereas
	       $l_{m,n}$ (relating with the wave number in the 1D quantum
	       system) are fixed. 

	 \item Correlation and scaling dimension for marginal operator
	       \begin{equation}
  \langle O_{marg} (z) O_{marg} (0) \rangle \propto |z|^{-4}.
   \label{eq:scaling-dim2}
	       \end{equation}
	       Thus, the scaling dimension for the marginal operator is 
 \fbox{$x_{marg}=2,l=0$} . 
	\end{enumerate}

  \item{Interacting case ($y_2 \neq 0$)}

	\begin{enumerate}
	 \item Parity ($\phi \rightarrow -\phi$; 
	       corresponding to the space inversion in 1D
	       quantum system)

	       Considering parity, we choose operators 
	       $ \cos(\sqrt{2} n \phi)$ and $\sin(\sqrt{2} n \phi) $.

	 \item{Renormalization from $y_1$ term}

	       Couplings $y_1$ and $y_2$ are renormalized as eqs. 
	       (\ref{eq:BKT-renormalization}).  
	       This affects all the scaling dimensions, 
	       according as eq. (\ref{eq:scaling-dim}).
	\end{enumerate}
 \end{enumerate}

 \subsection{Normal BKT ($p=1$)}

 In this case, the $y_2$ coupling term in (\ref{eq:sine-Gordon}) becomes
  relevant at $K=4$ on $y_{2}=0$.
 \begin{enumerate}
	 \item{Level crossing at $K=4, y_{2}=0$}

	       On Gaussian line at $K=4$, according to 
	      eqs. (\ref{eq:scaling-dim}), (\ref{eq:scaling-dim2}), 
	      five operators 
	      ($(m= \pm 4, P=1, q=0), (m=0, P=1, q=0), 
	      (m=0, P=-1, q=0), (m=0, P=1, q=0)$)
	       have the same scaling dimensions ($x=2$).

	 \item{Hybridization ($y_2 \neq 0$)}

	      The operator $\cos \sqrt{2} \phi$ and the marginal operator are
	      affected from
	      the renormalization of $y_2$, 
	       since they have the same symmetry ($m,P$),  
	       they hybridize each other by the $y_2$ term 
	      \cite{Nomura} (calculation can be more simplified
	      with the operator product expansion (OPE) \cite{Nomura-Kitazawa}). 
	      Combining the renormalization from $y_{1}$, there remains 
	      a level crossing on the BKT lines (see table \ref{table:table1}). 
 \end{enumerate}

 \subsection{BKT with $Z_2$ ($p=2$)}

 In this case, the $y_2$ coupling term in (\ref{eq:sine-Gordon}) becomes
  relevant at $K=1$ on $y_{2}=0$.

 \begin{enumerate}
	 \item{Level crossing at $K=1, y_{2}=0$}

	       On Gaussian line at $K=1$, besides
	      the level crossing at $x=2$, 
	      the four operators 
	      ($(m= \pm 1, P=-1, q=\pi), (m=0, P= \pm 1, q=\pi)$)
	       have the same scaling dimensions ($x=1/2$), 

	 \item{Level split ($y_{2} \neq 0$)}

	      With the $\cos 2 \sqrt{2} \phi$ coupling, scaling dimensions for
	      $\cos \sqrt{2} \phi$ and $\sin \sqrt{2} \phi$ are split 
	      \cite{Giamarchi-Schulz, Nomura-Okamoto}.
	      Combining the renormalization from $y_{1}$, 
	      there remains a level crossing on the BKT lines
	      (see table \ref{table:table2}). 
 \end{enumerate}

 \section{Physical examples}

Here we show physical examples for the quantum spin and the electron chains.  
Note that spin systems obey the commutation relations, whereas 
electron systems obey the anti-commutation relations.  
Thus, for the spin chain, one can directly relate the quantum numbers 
to those of the sine-Gordon model, whereas for the electron case, 
we should choose an appropriate boundary
condition according to evenness or oddness of quantum numbers (selection rule).

 \subsection{S=1 spin chain}
 First we consider the S=1 bond-alternating XXZ chain, 
 \begin{equation}
  H= \sum_{j=1}^{L} (1-\delta (-1)^j)
   (S^x_j S^x_{j+1} +S^y_j S^y_{j+1}+ \Delta S^z_j S^z_{j+1}) .
   \label{eq:bond-al-Hamil}
 \end{equation}
 In this case, quantum numbers are defined as the magnetization $m=\sum
 S^z_j$, the parity $P$ for the space inversion $S_j \leftrightarrow S_{L+1-j}$,
 the wave number $q$ for the translation by two sites $S_j \rightarrow S_{j+2}$.  
 The level crossing of excitations ($m=\pm 4, P=1,q=0$) and 
 ($m= 0, P=1,q=0$) corresponds to the BKT phase boundary.  
 The detailed analysis of the phase diagram and the universality class for this
 model is given in \cite{Kitazawa-Nomura-Okamoto, Kitazawa-Nomura}.  
 Note that using TBC method, one can improve the accuracy \cite{Kitazawa,
 Nomura-Kitazawa}.

 \subsection{S=1/2 spin chain}
 Next we consider the S=1/2 XXZ chain with the next-nearest-neighbor interaction.
 \begin{equation}
  H= \sum_{j=0}^{L-1} (h_{j,j+1} + \alpha h_{j,j+2}), \qquad
   h_{i,j}= S^x_i S^x_j +S^y_i S^y_j+ \Delta S^z_i S^z_j .
   \label{eq:NNN-Hamil}
 \end{equation}
 In this case, the level crossing of excitations ($m=\pm 1,q=\pi$) and 
 ($m= 0, P=\pm 1,q=\pi$) corresponds to the BKT phase boundary.  
 The detailed analysis of the phase diagram and the universality class for this
 model is given in \cite{Nomura-Okamoto}.  

 \subsection{Electron system: selection rule}

We briefly review the history of the selection rule and the boundary
condition of the 1D fermion model.  
Using the Jordan-Wigner (non-local) transformation, 
Lieb et al. \cite{Lieb-Schultz-Mattis} have studied the exact mapping
from a S=1/2 spin chain, which is equivalent to a hard core boson, 
to a spinless fermion chain.  They have pointed out that 
according to the oddness or evenness of the fermion number, 
one should use the PBC or TBC.  
In the sine-Gordon model (phase Hamiltonian) mapped from the spinless fermion,
or the bosonization language, Haldane \cite{Haldane} has written a
systematic review. 
In that paper, he has introduced a new quantum number, current,
$J=N_{L}-N_{R}$, and he has written a selection rule for fermion numbers
and current number. One can see from eq. (3.54) in
\cite{Haldane}, that boundary condition should change according to these
quantum numbers (although in \cite{Haldane} only the forward scattering
case was discussed, it is possible to include the umklapp
interacting case).  
The extension from the spinless fermion case to the electron chain
(fermion with the spin freedom) is straightforward,
considering two species of spinless fermion chains. 
From the another point of view, the selection rule has been found in the
field of Bethe Ansatz.  
For the Hubbard model, it is described by Woynarovich \cite{Woynarovich}.

Returning to the concrete procedure, 
we consider the spinless fermion case given by 
\begin{equation}
H = -t\sum_{j=1}^{L}\left[c_{j}^{\dagger}c_{j+1}+c_{j+1}^{\dagger}c_{j}\right]
+ V\sum_{j=1}^{L}c_{j}^{\dagger}c_{j}c_{j+1}^{\dagger}c_{j+1}.
\label{eq:t-V}
\end{equation}
For the ground state, we choose the following boundary condition. 
When the particle number is odd, we assume PBC. 
But when $N$ is even, the ground state is two-fold degenerate. 
In order to remove this degeneracy, we assume TBC $c_{j+L} = -c_{j}$. 
We write the fermion number for the left mover as $N_{L}$, 
and for the right mover as $N_{R}$
(we do not include $q=0$ fermion as $N_{L}$ or $N_{R}$). 
We define another numbers $m$ and $n$ as
\begin{equation}
  m = N_{L}+N_{R}-N_{0}, \hspace{5mm} n = \frac{N_{L} - N_{R}}{2}. 
\end{equation}
When the particle number is odd, $N_{0}$ is the particle number 
of the ground state minus 1 with PBC, and 
when the particle number is even, $N_{0}$ is the particle number 
of the ground state with TBC. 
$m$ means the change of the particle number. 
These numbers relate to the boundary condition of the phase field \cite{Haldane}
\begin{equation}
  \phi(x+L) = \phi(x) -\sqrt{2}\pi m, \hspace{5mm}
  \theta(x+L) = \theta(x) -\sqrt{2}\pi n.
\end{equation}
The low-lying excitation spectrum of the system is given by
\begin{equation}
  E_{m,n}(\tilde{n}_{L},\tilde{n}_{R})-E_{0} 
  = \frac{2\pi v}{L}x_{m,n}
  + \frac{2\pi v}{L} (\tilde{n}_{L}+\tilde{n}_{R})
\label{eq:ex-energy}
\end{equation}
where $E_{0}$ is the ground state energy, 
$v$ is the sound velocity, and $x_{m,n}$ is given by 
eq. (\ref{eq:scaling-dim}) which is the scaling dimension of the
operator (\ref{eq:vertex-operator}).
The second term of eq. (\ref{eq:ex-energy}) gives 
the sound wave collective excitation 
and $\tilde{n}_{L,R}$ are non-negative integers. 
The wave number of this excitation is given by
\begin{equation}
  q = -\left(2k_{F}+\frac{2\pi}{L}m\right)n
    -\frac{2\pi}{L}(\tilde{n}_{L} - \tilde{n}_{R}), 
\end{equation}
where $k_{F}$ is the Fermi wave number. 
Since $N_{L}$ and $N_{R}$ are integer, the number $n\pm m/2$ must be so. 
Thus {\em when $m$ is an even integer, $n$ is an integer, 
and when $m$ is an odd integer, $n$ is half odd integer}
($(-1)^{m} = (-1)^{2n}$).  
For Tomonaga and Luttinger models, $N_{L}$ and $N_{R}$ are good quantum 
numbers. When some interactions which do not conserve $N_{L}$ and $N_{R}$, 
such as the umklapp scattering, are introduced, 
$m$ remains a good quantum number but $2n$ does not. 
{\em
Although only the parity $(-1)^{2n}$ is conserved, 
it does not violate the selection rule mentioned above. }
The instability of the BKT transition for the model (\ref{eq:t-V}) 
occurs at $V=2t$ 
for the half filling case $\sum_{j}c_{j}^{\dagger}c_{j} = L/2$, 
and this instability stems from the umklapp scattering. 

Next we consider the electron case with spin freedom, such as the 
1D extended Hubbard model
\begin{eqnarray}
  H &=& -t\sum_{j=1}^{L}\sum_{\sigma=\uparrow\downarrow}
  \left[c_{j\sigma}^{\dagger}c_{j+1\sigma}+c_{j+1\sigma}^{\dagger}c_{j\sigma}
  \right] 
  + U\sum_{j=1}^{L}n_{j\uparrow}n_{j\downarrow}
\nonumber \\
  && + V\sum_{j=1}^{L}
    \left(n_{j\uparrow}+n_{j\downarrow}\right)
    \left(n_{j+1\uparrow}+n_{j+1\downarrow}\right),  
\end{eqnarray}
where $n_{j\sigma} = c_{j\sigma}^{\dagger}c_{j\sigma}$. 
In this case, we have four particle numbers of the 
left and the right movers with spin $\sigma = \uparrow\downarrow$, 
$N_{L\sigma}$ and $N_{R\sigma}$. 
We assume PBC when the particle number of the ground state is $4M+2$ 
($M$ is integer), and TBC when $4M$. 
In both cases, we define $N_{0} = 4M$. 
When the system has an SU(2) symmetry relating to the spin freedom, 
the Fermi wave numbers for $\sigma =\uparrow$ and $\downarrow$ are same 
$k_{F\uparrow} = k_{F\downarrow} = k_{F}$, and we define the new numbers 
\cite{ Woynarovich, Nakamura98}
\begin{eqnarray}
m_{c} &=& (N_{L\uparrow} + N_{R\uparrow}+N_{L\downarrow} + N_{R\downarrow})/2
  -N_{0}/2,
  \nonumber \\
n_{c} &=& (N_{L\uparrow} - N_{R\uparrow}+N_{L\downarrow} - N_{R\downarrow})/2,
  \nonumber \\
m_{s} &=& (N_{L\uparrow} + N_{R\uparrow}-N_{L\downarrow} - N_{R\downarrow})/2,
\label{eq:numbers} \\
n_{s} &=& (N_{L\uparrow} - N_{R\uparrow}-N_{L\downarrow} + N_{R\downarrow})/2.
  \nonumber 
\end{eqnarray}
$2m_{c}$ gives the total change of the electron number.
The boundary condition of the phase field is given by 
\begin{equation}
  \phi_{c,s}(x+L) = \phi_{c,s}(x) - \sqrt{2}\pi m_{c,s},
\hspace{5mm}\theta_{c,s}(x+L) = \theta_{c,s}(x) - \sqrt{2}\pi n_{c,s}. 
\end{equation}
Numbers in eq. (\ref{eq:numbers}) relate to the energy spectrum as 
\begin{eqnarray}
  \lefteqn{ E_{m_{c},n_{c},m_{s},n_{s}}
  (\tilde{n}_{Lc},\tilde{n}_{Rc},\tilde{n}_{Ls},\tilde{n}_{Rs}) - E_{0} }
\nonumber \\
  &=& 
  \frac{2\pi v_{c}}{L}\left(\frac{1}{2K_{c}}m_{c}^{2} 
  + \frac{K_{c}}{2}n_{c}^{2}\right)
  +\frac{2\pi v_{s}}{L}\left(\frac{1}{2K_{s}}m_{s}^{2} 
  + \frac{K_{s}}{2}n_{s}^{2}\right)
\\
  && + \frac{2\pi v_{c}}{L}(\tilde{n}_{cL} + \tilde{n}_{cR})
  + \frac{2\pi v_{s}}{L}(\tilde{n}_{sL}+\tilde{n}_{sR}),
\nonumber
\end{eqnarray}
with the wave number 
\begin{equation}
  q
= -\left(2k_{F}+\frac{2\pi}{L}m_{c}\right)n_{c} 
  -\frac{2\pi}{L}m_{s}n_{s} 
  -\frac{2\pi}{L}(\tilde{n}_{cL} - \tilde{n}_{cR}
  +\tilde{n}_{sL} - \tilde{n}_{sR}). 
\end{equation}
Relating to the SU(2) symmetry of the system, we have $K_{s}=1$. 
Integer numbers $N_{L\sigma}$ and $N_{R\sigma}$ give the 
selection rule for the excitation: 
{\em When $2m_{c}$ is even (odd) integer, $n_{c}$, $m_{s}$, and $n_{s}$ 
are integer (half odd integer). 
And when $m_{c}+m_{s}$ is even (odd) integer, $n_{c}+n_{s}$ is even (odd). }

The level spectroscopy with the selection rule has been applied for the
spin-gap problem of
the $t-J$ model \cite{Nakamura-Nomura-Kitazawa,Nakamura98}, and for the extended Hubbard
model (half and quarter-filling) \cite{Nakamura}. 

\bigskip

For other applications, {\em e.g.}, magnetic plateau, spin-Peierls
transition, please see references in \cite{Okamoto-2002}

\section{Acknowledgement (en francais)}

Premi\`{e}rement, nous remercions Dr. K. Okamoto (application aux plateaux
magnetiques) et M. Nakamura (application aux syst\`{e}mes \'{e}lectroniques).

Cette th\'{e}orie \'{e}tait
 n\'ee, quand un des auteurs, K. N., etudiait en France, il y a sept ans. 
Le nom ``level spectroscopy'' \'{e}tait nomm\'{e} 
par professeur H. J. Schulz.
Nous remercions profondement H. J. Schulz, parce qu'il a encourag\'e
 K. N. \`{a}
d\'{e}velopper cette id\'ee, et qu'il a appr\'eci\'e cette th\'{e}orie,
apr\`{e}s l'accomplissement. 
L'embryon de cette id\'ee remonte aux recherches
\cite{Ziman-Schulz, Affleck-Gepner-Schulz-Ziman}

Indirectement, les discussions sur la correction logarithmique 
avec professeur I. Affleck par lettres (e-mail) ont aid\'e le
progr\`{e}s de notre recherche. 
Professeur M. Takahashi nous a invit\'{e} \`{a} commencer l'\'etude sur la
theorie des champs conforme. 
Les crititiques et les commentaires par professeur H. Shiba, par
example,
``C'est tr\`{e}s difficile \`{a} rechercher num\'{e}riquement la supraconduction,
parce il y a la probl\`{e}me de renormalization.'', aussi sont de valeur.

\end{document}

%% file: table1.tex
 \begin{tabular}{|ccc||c||l||l|l|}
  \hline
  m       & P & q & BC & $x$ & operator in s.G. & abbr. \\
  \hline
  $\pm 2$ & 1 & 0 & PBC & $ 1/2 - y_1(l)/4$ & $\exp(\pm i 2 \sqrt{2} \theta )$  & $x_{\pm 2,0}$ \\
  0       & $-1^{*}$ &  & TBC & $1/2 + y_1(l)/4 - y_2(l)/2$ & $\sin (\phi / \sqrt{2})$  & $x_{0,sin}^{TBC}$ \\
  0       & $1^{*}$  &  & TBC & $1/2 + y_1(l)/4 + y_2(l)/2$ & $\cos (\phi / \sqrt{2})$ & $x_{0,cos}^{TBC}$\\
  \hline
  $\pm 4$ & 1 & 0 & PBC & $2 - y_1(l)$ & $\exp(\pm i 4 \sqrt{2} \theta )$ &
$x_{\pm 4,0}$ \\
  0       & 1 & 0 & PBC & $2 - y_1(l )(1 + 4t/3)$ & marginal & $x_{marg}$\\
  0       & -1& 0 & PBC & $2 + y_1(l)$ & $\sin (\sqrt{2} \phi )$ &$x_{0,sin}$ \\
  0       & 1 & 0 & PBC & $2 +  2 y_1(l)(1 + 2t/3)$ & $\cos (\sqrt{2} \phi )$ & $x_{0,cos}$\\
  \hline
 \end{tabular}

%% file: table2.tex
 \begin{tabular}{|ccc||l||l|l|}
  \hline
  m       & P  & q & $x$ & operator in s.G. & abbr.\\
  \hline
  $\pm 1$ &    & &$1/2 - y_1(l)/4$ & $\exp(\pm i \sqrt{2} \theta )$ & $x_{\pm 1,0}$ \\
  0       & -1 & $\pi$ & $1/2 + y_1(l)/4 - y_2(l)/2$ & $\sin (\sqrt{2} \phi )$  & $x_{0,sin}$ \\
  0       & 1  & $\pi$ & $1/2 +  y_1(l)/4 + y_2(l)/2$ & $\cos (\sqrt{2} \phi )$ & $x_{0,cos}$ \\
  \hline
  $\pm 2$ & 1  & 0 & $2 - y_1(l)$ & $\exp(\pm i 2 \sqrt{2} \theta)$ & $x_{\pm 2,0}$ \\
  0       & 1  & 0 & $2 - y_1(l)(1 + 4t/3)$ & marginal & $x_{marg}$ \\
  0       & -1 & 0 & $2 + y_1(l)$ & $\sin ( 2 \sqrt{2} \phi )$ & $x_{0,sin2}$ \\
  0       & 1  & 0 & $2 + 2 y_1(l)(1 + 2t/3)$ & $\cos ( 2 \sqrt{2} \phi )$ & $x_{0,cos2}$ \\
  \hline
 \end{tabular}